\documentclass[aps,twocolumn,prl,showpacs]{revtex4}
\usepackage{amsmath}
\usepackage{amssymb}
\usepackage{graphicx}
\usepackage{bm}
\usepackage{longtable}

\begin{document}
\title{Magnetic anisotropy of singly Mn-doped InAs/GaAs quantum dots}

\author{Olivier Krebs}
\author{Emile Benjamin}
\author{Aristide Lema\^{i}tre}
\affiliation{Laboratoire de Photonique et Nanostructures-CNRS, Route de Nozay, 91460
Marcoussis, France}

\def\xp{$X^{+}$}
\def\xm{$X^{-}$}
\def\x0{$X^{0}$}
\def\xx0{$2X^{0}$}
\def\etal{~\textit{et al.}~}
\date{\today}

\begin{abstract}

We report on the micro-photoluminescence  spectroscopy of InAs/GaAs quantum dots (QD)  doped by a single Mn atom in a magnetic field either  longitudinal or perpendicular to the optical axis. In both cases the  spectral features of  positive trion ($X^+$)  are found to split into strongly circularly  polarized components, an effect  very surprising  in a perpendicular magnetic field. The field-induced splitting is ascribed to the transverse Zeeman splitting of the  neutral acceptor complex $A^0$ issued by the Mn impurity, whereas the  circular optical selection rules  result from the $p$-$d$ exchange which acts as a very strong longitudinal magnetic field inhibiting the spin mixing by the transverse field of the QD heavy-hole ground state.  A theoretical  model of the spin interactions which includes (i) the local strain anisotropy  experienced by the acceptor level and (ii) the anisotropic exchange due to the out-of-center Mn position  provides a very good agreement with our observations.
\end{abstract}

\maketitle

\indent Doping a semiconductor quantum dot (QD) with a single Mn atom brings up remarkable spin-related properties due to the sp-d exchange interactions between the confined carriers (electron and hole) and the magnetic impurity. In the last few years Mn-doped CdTe QD's have been extensively studied  by micro-photoluminescence ($\mu$-PL) spectroscopy in an  external magnetic field\cite{Besombes04,Besombes05,Leger06,Leger05a}. Most of the observations were very well interpreted by assuming a 5/2 spin for the Mn ion acting on the carriers confined in a quantum dot through Heisenberg Hamiltonians. Yet, the strong vertical confinement of QD's along their growth axis, as well as their in-plane biaxial strain were shown to modify significantly the spectral features because of  the resulting heavy-hole nature of the valence band ground state.\\
\indent The quite specific signature of  InAs/GaAs quantum dots doped with a single Mn atom  has been recently uncovered in  $\rm\mu$-PL spectroscopy~\cite{Kudelski07}. In this system, the Mn impurity acts as an effective $J=1$ spin with   a noticeable fine structure splitting in zero magnetic field. This results from the  neutral acceptor  ($A^0$) complex formed by Mn in a III-V matrix, namely a negatively charged center $A^-$  and a bound hole $h_1$~\cite{Szczytko96,Schneider87,Bhattacharjee99,Govorov04}.  The $J=1$ spin corresponds to the ground state of the 3$d^5$ Mn spin $S=5/2$ and the bound hole total angular momentum $J_{h_1}=3/2$ which interact via the anti-ferromagnetic $p$-$d$ exchange. Its zero-field splitting  results from some local anisotropy of the  potential experienced by the bound hole\cite{PRBvanBree08,Kudelski07,CRP-Govorov08}. Within this interpretation, the  anisotropy of the $A^0$ complex does not affect  the optical selection rules of the QD interband transitions which still involve a conduction  electron ($e$)  and a valence band hole  ($h_2$)  essentially of  heavy-hole character both with  $S$-like orbital. This was shown in Ref.~\onlinecite{Kudelski07} where a longitudinal magnetic field split all the optical transitions into their circularly polarized ($\sigma^\pm$) components.  In this article, we show  that the optical selection rules in a transverse magnetic field are in contrast deeply affected by the anisotropy of the $A^0$ effective spin, besides  in a rather non-intuitive way. Indeed, in  a magnetic field perpendicular to the optical axis, the optical transitions which are expected to be linearly polarized as  usually encountered in undoped InAs QDs~\cite{PRB-Leger07,PRL-Xu07,PRB-Koudinov04},  exhibit for Mn-doped InAs QDs a strong circular polarization ($\sigma^+$ or $\sigma^-$). We show that this effect results from the $A^0$ spin anisotropy  which enables to split by Zeeman effect the ferromagnetic (FM) and anti-ferromagnetic (AFM) configurations of the  $h_2$-$A^0$ complex, while the heavy-hole $h_2$  keeps a well-defined pseudo-spin $\Uparrow$ ($J_{h_2,z}=+3/2$) or $\Downarrow$ ($J_{h_2,z}=-3/2$).\\
\indent We studied a sample grown by molecular beam epitaxy on a semi-insulating GaAs [001] substrate which  consists of a single layer of InAs/GaAs QD's   randomly doped by a single Mn atom  (see Ref.~\onlinecite{Kudelski07} for details). We estimate that $\sim$0.1 to  1\% of the quantum dots are effectively doped by a single Mn atom. The $\mu$-PL spectroscopy of individual Mn-doped InAs QDs was carried out with a split-coil magneto-optic cryostat. A 2~mm focal length aspheric lens (N.A.~0.5) actuated by piezo motors was used to focus the He-Ne excitation laser and to collect the PL from the sample. This  compact microscope, which integrates both the sample and the optical lens, can be rotated about the vertical axis of the cryostat in order to change  the magnetic field direction with respect to the optical axis from parallel (Faraday configuration) to perpendicular (Voigt configuration). Relying on in-situ  sample imaging  we could  therefore study the same quantum dot in both configurations.  All measurements presented here were performed at low temperature (T=2~K). The  collected PL was dispersed by a 0.6~m-focal length double spectrometer and detected by a Nitrogen-cooled CCD array camera.\\
\begin{figure}[h] \includegraphics[width=0.48 \textwidth,angle=0]{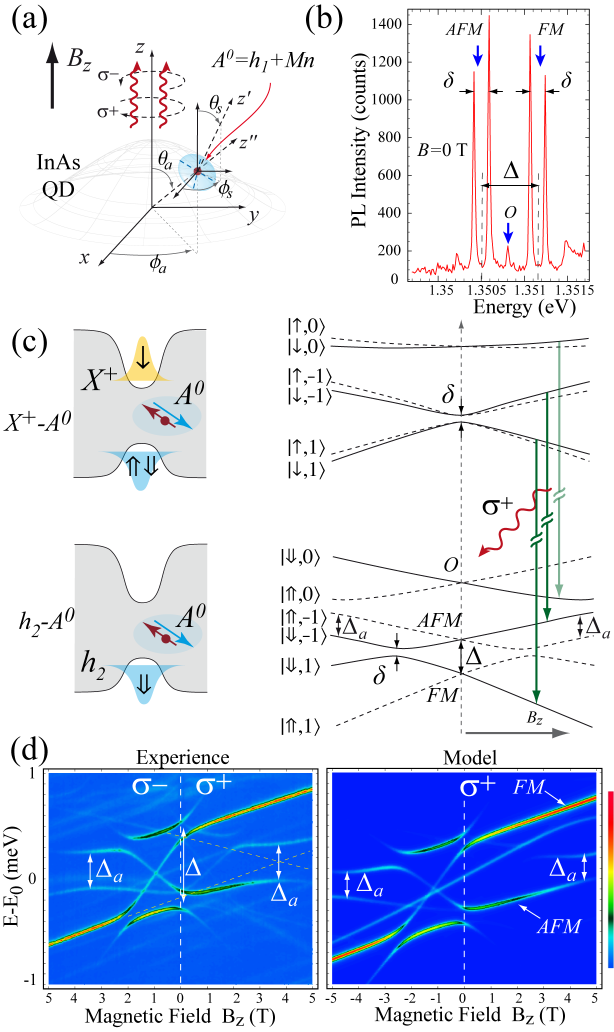} \caption{(Color Online) QD1 (a) Schematics of an InAs QD doped by a single Mn atom forming a neutral acceptor $A^0$. (b) PL spectrum of a charged exciton $X^+$ in such a QD at zero magnetic field. (c) Diagram of  energy levels involved in the  $X^+$-$A^0\rightarrow h_2$-$A^0$ transition. Solid (dashed) lines correspond to the levels with $\downarrow$ ($\uparrow$) $e$ spin and $\Downarrow$ ($\Uparrow$) $h_2$ pseudo-spin connected via a $\sigma^+$ ($\sigma^-$) photon. (d) Contour-plot of QD1 PL around $E_0=1.355$~eV as a function of  longitudinal magnetic field  measured in $\sigma^-$  and $\sigma^+$ circular polarization (left) and  theoretical simulation (right). The diamagnetic shift is subtracted so that the linear dependence on $B_z$ due to the Zeeman effect produces straight  lines.} \label{Fig1} \end{figure}
\indent Figures \ref{Fig1} and \ref{Fig2} report on the optical spectroscopy of a charged exciton $X^+$ in the same Mn-doped InAs QD (QD1) measured respectively in Faraday and Voigt configuration. Let us first comment  on the results shown in Fig.~\ref{Fig1}. As evidenced in Ref.~\cite{Kudelski07}, the Mn-doped InAs QD is identified by its spectral features in zero magnetic field shown in Fig.~\ref{Fig1}(b). It consists of two doublets separated by the  exchange energy $\Delta$ between the FM and AFM configurations of $h_2$-$A^0$, plus a weaker line denoted $O$ which  corresponds to the transition involving the $A^0$ state $J_z=0$. Another  specific feature is the equal splitting  $\delta$ of both FM and AFM doublets which is ascribed to the fine structure of $A^0$ in its anisotropic environment.\\
\indent When the magnetic field is applied parallel to the optical axis $z$ the quantum levels involved in these optical transitions are split due to the Zeeman effect (see Fig.~\ref{Fig1}(c)), which gives rise to a very distinctive contour-plot of the PL intensity as shown in Fig.~\ref{Fig1}(d). This 'magneto-PL' image is  composed of a series of  spectra measured in both $\sigma^+$ and $\sigma^-$ polarizations, while the magnetic field was changed step by step with an increment of 50~mT from 0 to 5~T. The PL intensity is plotted against a linear color scale, and interpolation was used for a better graphical rendering.  We observe a characteristic cross-like pattern, quite similar to the one reported in Ref.~\onlinecite{Kudelski07} for a negatively charged exciton ($X^-$), with yet a noticeable difference. Here the cross  shows up in negative fields (i.e. in $\sigma^-$ polarization)  while it was previously found  in positive fields. This kind of mirror symmetry is similar to that observed in Refs.~\onlinecite{Kudelski07,Besombes05} between an exciton and a biexciton.  Here, it corresponds to the symmetry between a positive trion ($X^+$)  and a negative trion ($X^-$). Indeed, both involve the same spin configurations either in the final or initial state of the optical recombination with yet orthogonal selection rules due to the Pauli principle. (e.g. for $X^+$ $|\Uparrow\Downarrow\uparrow\rangle\stackrel{\sigma-}{\overrightarrow{\qquad}}|\Uparrow \rangle$ while for $X^-$ $|\uparrow\downarrow\Uparrow\rangle\stackrel{\sigma+}{\overrightarrow{\qquad}}|\uparrow \rangle$). Note that  in this  sample most of the quantum dots (with or without Mn doping)  were found as positively charged by an excess hole, after measuring the sign of the Overhauser shift (namely the sign of nuclear polarization) generated under quasi-resonant circularly polarized excitation\cite{PRB-Eble,CRP-Krebs}.\\
\indent The $A^0$ exchange anisotropy  originates from the position of the Mn atom out of the QD center. One can  distinguish two main contributions: (i) the anisotropy of the local potential experienced by the $h_1$ bound hole which lifts the degeneracy of the $J=1$ state of $A^0$~\cite{Kudelski07,Govorov04,Schneider87}, (ii) the anisotropic part of the exchange coupling between  the $h_2$ hole and the out-of-center Mn spin~\cite{PRB-Bhattacharjee07}. The first contribution is responsible for the coupling  of the $|J_z=\pm 1\rangle$ $A^0$ states which gives rise  to their  anticrossing $\delta$ when they are brought into coincidence. This occurs in zero magnetic field for the $X^+$-$A^0\equiv e$-$A^0$ levels  because the effective  exchange with $X^+$ (namely with a single $e$ spin) is essentially zero  here, and at $B_z\approx\pm$2~T for the $h_2$-$A^0$ levels (see Fig.~\ref{Fig1}(c)). More generally, this anisotropy is also responsible  for the forbidden transitions corresponding to $|\Delta J_z|=2$ which form the cross pattern and which are partially permitted because  $J_z$ is  not a good quantum number. The second contribution is responsible for the weaker lines  in Fig.~\ref{Fig1}(d) which gives rise to the anticrossings denoted $\Delta_a$. As discussed below, they result from a mixing of the $|\Uparrow,\pm1\rangle$ and $|\Downarrow,\pm1\rangle$ $h_2$-$A^0$ states which makes visible the normally forbidden transitions $|\uparrow,\pm1\rangle \rightarrow|\Downarrow,\pm1\rangle$. Note however that all the lines keep a very strong circular polarization, because the  $e$-$A^0$  initial states still have  pure electron spin $\uparrow$ or $\downarrow$.\\
\indent In Voigt configuration, the $A^0$ exchange anisotropy  gives rise to a very surprising signature. When the transverse magnetic field is increased each line originating from the zero field  doublets  splits into four lines with a strong $\sigma^+$ or $\sigma^-$  circular polarization as reported in Fig.~\ref{Fig2}(b). Such a polarization is quite unexpected because, due to the mixing of the spin states for both the electron and hole by the  field, the optical transitions of  trions should become linearly polarized as observed in non-magnetic QDs\cite{PRB-Leger07,PRL-Xu07,PRB-Koudinov04,Fernandez-PRL09}. Note the time reversal symmetry is yet well respected here since changing the sign of the applied magnetic field reverses the circular polarization from $\sigma^\pm$ to $\sigma^\mp$ (see Fig.~\ref{Fig2}(a)). This clearly implies  that the PL spectra  depend on the magnetic field direction in the $xy$ plane because changing the sign of $B_x$ can be achieved by a $\pi$-rotation about $z$, which points out the  role of $A^0$ in-plane anisotropy.\\
\indent We have modeled these  magneto-PL images by considering  the  spin Hamiltonians of the four involved particles (Mn, $h_1$, $e$, $h_2$) assumed to occupy the ground state of their respective confinement potential. $e$ and $h_2$  have essentially the $S$-like character of the respective conduction and valence QD ground sate, while  $h_1$  is assumed to be strongly localized onto the Mn site which lies itself at a certain distance from the QD center (see Fig.~\ref{Fig1}(a)). The theoretical indiscernibility between both overlapping holes  is phenomenologically treated by introducing an exchange hamiltonian between their spins (see below), which amounts to treat this issue in the Heitler-London scheme~\cite{Fazekas}. This approach enables us to restrict ourselves to the spin degree of freedom of the four involved particles ($S_\text{Mn}=5/2$, $J_{h_1}=3/2$, $S_e=1/2$ and  $J_{h_2}=3/2$),  with interactions described using only their respective spin operators.  The single-particle Hamiltonians read: \begin{eqnarray} \hat{\mathcal{H}}_{\text{Mn}}&=&g_{\text{Mn}}\, \mu_\text{B}\, \hat{\bm{S}}_{\text{Mn}}\cdot\bm{B}\nonumber\\ \label{eq:Zeeman} \hat{\mathcal{H}}_{h_1} &=& g_{h_1}\, \mu_\text{B}\, \hat{\bm{J}}_{h_1}\cdot\bm{B}+\hat{\mathcal{H}}_s\\ \hat{\mathcal{H}}_{h_2}&=&g_{h}\,\mu_\text{B}\,  \hat{\bm{J}}_{h_2}\cdot\bm{B}+\hat{\mathcal{H}}_{HL}\nonumber \\ \hat{\mathcal{H}}_{e}&=&g_{e}\, \mu_\text{B}\, \hat{\bm{S}}_e\cdot\bm{B} \nonumber \end{eqnarray} where $g_{\alpha}$ denotes the Land\'{e} factor of the  particle $\alpha$ taken to the first order as a scalar for Mn, $h_1$ and $h_2$, while for $e$  a longitudinal ($g_{e,z}$) and transverse ($g_{e,\perp}$) factor will be used. In this formalism, the well-known anisotropy of the Land\'{e} factor for the  hole states will naturally result from their  splitting into  heavy- and light-hole states. The  potential anisotropy $\hat{\mathcal{H}}_s$ experienced by  $h_1$ can be described via an effective strain tensor with three main axes that in general differ from the crystallographic axes\cite{Yakunin07}. In that aim, we introduced as the dominant term a compressive strain $\epsilon_{\parallel}$ along  the $z'$ direction defined by two angles $\theta_s$ and $\phi_s$  (see Fig.~\ref{Fig1}(a)) plus an in-plane shear strain $\epsilon_{\perp}$ with $x'$ and $y'$ main axes defined by an angle $\psi_s$ in the plane perpendicular to $z'$. $\hat{\mathcal{H}}_s$ reads then:
 \begin{eqnarray}
\hat{\mathcal{H}}_s&=&- \frac{\epsilon_{\parallel}}{3}\left[\hat{J}^2_{h_1,z'}-\frac{1}{2}\left( \hat{J}^2_{h_1,x'}+\hat{J}^2_{h_1,y'}\right)\right]\nonumber\\
& &\qquad \qquad \qquad \qquad + \frac{\epsilon_{\perp}}{2} \left( \hat{J}^2_{h_1,x'}-\hat{J}^2_{h_1,y'}\right) \label{eq:Hs}
\end{eqnarray}
 where the $\hat{J}^2_{h_1,\alpha'}$'s deduce from the $\hat{J}^2_{h_1,\alpha}$ operators by the three successive rotations, namely via the passage matrix $e^{-i \psi_s\hat{J}_{h_1,z}}\, e^{-i \theta_s\hat{J}_{h_1,y}}\, e^{-i \phi_s\hat{J}_{h_1,z}}$. For the  $h_2$ hole confined in the QD, we used a similar description, with $\Delta_{HL}\sim 30$~meV  the splitting between the heavy-hole and light-hole states due to the stronger confinement along $z$. The corresponding term reads:
 \begin{equation}
\hat{\mathcal{H}}_{HL}=- \frac{\Delta_{HL}}{3}\left[\hat{J}^2_{h_2,z}-\frac{1}{2}\left(\hat{J}^2_{h_2,x}+\hat{J}^2_{h_2,y}\right)\right]\label{eq:HHL}
\end{equation}
\\
\indent We assumed that the  exchange interactions between the different spins  take the form of Heisenberg Hamiltonians\cite{Bhattacharjee99,CRP-Govorov08}, with besides a specific anisotropic part $\mathcal{H}_a$ for the \emph{p-d} exchange between the out-of-center Mn and $h_2$\cite{PRB-Bhattacharjee07}. Since the $e$-$A^0$ exchange turns out to be negligible in our experimental observations, which could be due to weak overlap with the $A^0$ impurity, we only consider here the exchange interactions involving the Mn, $h_1$ and $h_2$  spins. The corresponding exchange Hamiltonian reads :
\begin{eqnarray}
\hat{\mathcal{H}}_{X} &=& \varepsilon_{\text{Mn}\text{-}h_1}\, \bm{\hat{S}}_{\text{Mn}}\cdot\bm{\hat{J}}_{h_1}+\varepsilon_{\text{Mn-}h_2} \bm{\hat{S}}_{\text{Mn}}\cdot\bm{\hat{J}}_{h_2}+\nonumber\\
& & \qquad\qquad\qquad\qquad+\varepsilon_{h_1\text{-}h_2} \bm{\hat{J}}_{h_1}\cdot\bm{\hat{J}}_{h_2} +\hat{\mathcal{H}}_a \quad\label{eq:HX}
\end{eqnarray}
where the   2-spin exchange energies $ \varepsilon_{\alpha\text{-}\alpha'}$ are considered as fitting parameters since they depend  on the actual overlap between the particles. Yet, we expect the Mn-hole exchange interaction to be anti-ferromagnetic as usually reported in literature. 
The anisotropic part $\hat{\mathcal{H}}_a$ in Eq.~\ref{eq:HX}  has been derived in Ref.~\onlinecite{PRB-Bhattacharjee07} in the case of a spherical quantum dot. To the first order, it depends linearly on a  parameter $\rho$ which depends itself on the Mn position and  vanishes when it lies at the QD center. Although our InAs quantum dots are  lens-shaped with no well-defined center, we assumed here the same expression for $\hat{\mathcal{H}}_a$ as follows:
\begin{equation}
\hat{\mathcal{H}}_a=\rho\,\, \varepsilon_{\text{Mn-}h_2}\,\left(\hat{J}_{h_2,z''}^2\bm{\hat{S}}_{\text{Mn}}\cdot\bm{\hat{J}}_{h_2}+\bm{\hat{S}}_{\text{Mn}}\cdot\bm{\hat{J}}_{h_2} \hat{J}_{h_2,z''}^2\right)
\label{eq:Ha}
\end{equation}\\
where the angular momentum operator $\hat{J}_{h_2,z''}$ along the direction $z''$ (defined by two angles $\theta_a$ and $\phi_a$ , see Fig.\ref{Fig1}(a))) refers to the position of the Mn atom with respect to some effective QD center, e.g.  the maximum of the $h_2$ $S$-like envelope function.\\
\indent To calculate the theoretical PL emission spectra, the Hamiltonians of the initial  and final states in \xp transition are first diagonalized. In practice we restricted ourselves to the first $A^0$ levels $J\!=\!1$ (as represented in Fig.~\ref{Fig1}(c)) and $J\!=\!2$ where $\bm{\hat{J}}=\bm{\hat{S}}_{\text{Mn}}+\bm{\hat{J}}_{h_1}$. The PL intensity   emitted in the transition from  $|i\rangle$ to  $|f\rangle$ is calculated by taking into account both the oscillator strength $\propto| \langle f|\hat{P}_{\sigma^+}|i\rangle|^2$ (where $\hat{P}_{\sigma^+}$ is the dipolar operator for a $\sigma^+$ polarization)  and the population $\rho_{ii}$ of the initial state $|i\rangle$. As  He-Ne excitation produces non-polarized carriers and electron spin relaxation may be considered negligible in QDs during the radiative lifetime of \xp ($\approx 1$~ns), the $\rho_{ii}$'s are determined only by the populations in the different states of $A^0$. Assuming that under weak optical excitation density,  thermalization occurs mostly in the final state (FS) of the system (here the $h_2$-$A^0$ complex) to a certain equilibrium  temperature $T_{\text{FS}}$, we deduced  the $\rho_{ii}$'s from the partial trace of the FS density matrix $\rho_{\text{\text{FS}}}= \exp(-\hat{H}_{\text{FS}}/k_\text{B}T_{\text{FS}})/ Tr[\exp(-\hat{H}_{\text{FS}}/k_\text{B}T_{\text{FS}})]$ where $k_\text{B}$ is the Boltzmann factor. Eventually, to compare with the experiments we applied to the calculated transitions a  Lorentzian broadening of FWHM=25~$\mu$eV. \\
\begin{figure}[h] \includegraphics[width=0.48 \textwidth,angle=0]{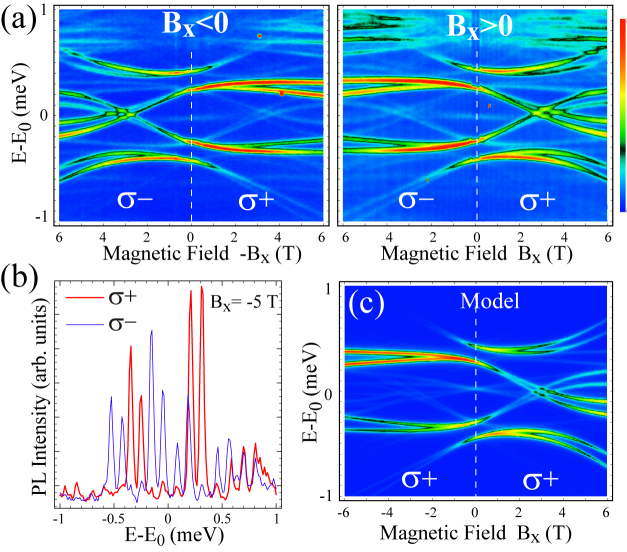} \caption{(Color online) QD1. (a) PL Contour-plot of the same \xp of Fig.~\ref{Fig1} as a function of   transverse magnetic field  $B_x$ positive (right) or negative (left). In both cases, the PL has been measured in $\sigma^-$  and $\sigma^+$ circular polarizations as indicated. The diamagnetic shift (half of that in Faraday configuration) has been subtracted. (b) PL spectra at $B_x$=-5~T in both $\sigma^+$ and $\sigma^-$ polarizations. (c) Theoretical simulation of the PL contour-plot (see text) with the same parameters as in Fig.~\ref{Fig1}. } \label{Fig2} \end{figure}\\

\begin{table}[h!]
\caption{Parameters used for theoretical simulations. Strains ($\epsilon_\parallel$,$\epsilon_\perp$) and exchange energies ($\varepsilon_{\text{Mn}\text{-}h_1}$, $\varepsilon_{\text{Mn-}h_2}$,  $\varepsilon_{h_1\text{-}h_2}$) are given in meV.}
\label{ParamTable}
\begin{ruledtabular}
\begin{tabular}{c | c c c c c c}
& $g_{Mn}$ & $g_{h_1}$ & $g_{h_2}$ & $g_{e,z}$ & $g_{e,\perp}$ &  $T_{\text{FS}} (K)$\\
& 2 & 0.8 & 0.85 & -0.6 & -0.35 &  15\\
\hline
\hline
&  $\epsilon_\parallel$ & $\epsilon_\perp$  & $\theta_s$  & $\phi_s$  & $\psi_s$ & \\
QD1  &   6.95 & 2.27 & 36$^\circ$ & -120$^\circ$ & 49$^\circ$ & \\
QD2  &   6.75 & 2.55 & 35$^\circ$ & -93$^\circ$ & 55$^\circ$ & \\
QD3  &   3.6  & 2    & 50$^\circ$ &  55$^\circ$ & 40$^\circ$ & \\
\hline
\hline
 & $\varepsilon_{\text{Mn}\text{-}h_1}$ & $\varepsilon_{\text{Mn-}h_2}$   & $\varepsilon_{h_1\text{-}h_2}$   & $\rho$ & $\theta_a$ & $\phi_a$ \\
QD1  & 4.2 & 0.6 & 1.2 & -0.1 & 81$^\circ$ & -120$^\circ$ \\
QD2  & 4.5 & 0.63 & 1.35 & -0.19 & 85$^\circ$ & -123$^\circ$ \\
QD3  & 5 & 0.85 & 1.8 & -0.12 & 102$^\circ$ & 55$^\circ$ \\
 \end{tabular}
 \end{ruledtabular}
\end{table}

\begin{figure}[h] \includegraphics[width=0.48 \textwidth,angle=0]{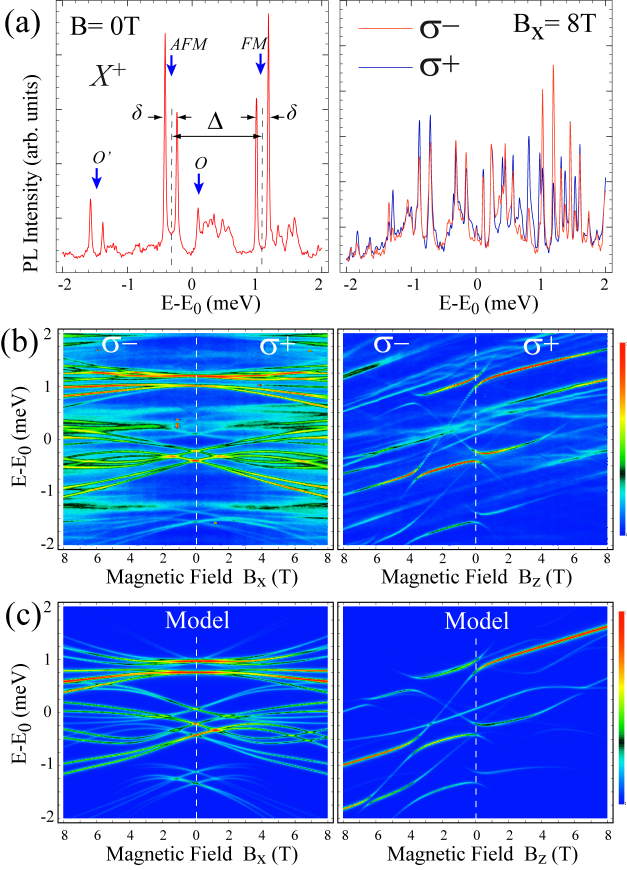} \caption{(Color online)  QD2. (a) PL spectra of an $X^+$ trion at $B$=0~T and $B_x$=8~T centered at $E_0=1.355$~eV. (b) PL Contour-plot as a function of   transverse (left) and longitudinal (right) magnetic field in $\sigma^-$  and $\sigma^+$ circular polarizations as indicated. The diamagnetic shift  is  subtracted. Additional  spectral features likely due to  different charge states of QD2 are also visible (c) Theoretical simulation of the PL contour-plot.} \label{Fig3} \end{figure}

\begin{figure}[h] \includegraphics[width=0.48 \textwidth,angle=0]{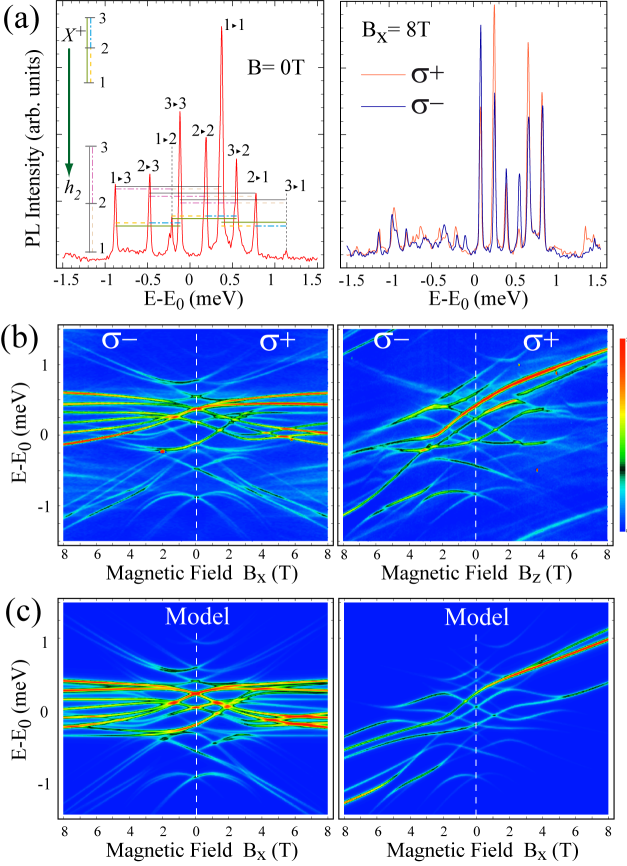} \caption{(Color online) QD3. (a) PL spectra of an $X^+$ trion at $B$=0~T and $B_x$=8~T centered at $E_0=1.3525$~eV. Inset at zero field shows the  3$\times$3 level fan chart responsible for the observed line splittings. (b) PL Contour-plot as a function of   transverse (left) and longitudinal (right) magnetic field in $\sigma^-$  and $\sigma^+$ circular polarizations as indicated. The diamagnetic shift  is  subtracted. (c) Theoretical simulation of the PL contour-plot.} \label{Fig4} \end{figure}

\indent The above model enabled us to reproduce quite well both magneto-PL images performed in Faraday and Voigt configurations  by using the same set of parameters (see Tab.~\ref{ParamTable}).  The electron and hole Land\'{e} factors were  determined  in order to reproduce the linear slopes due to the Zeeman effect. We found very similar values for  the three quantum dots investigated    here. Remarkably, thanks to the strong exchange field experienced by $h_2$, the transverse electron Land\'{e} factor $g_{e,\perp}$ could be determined unambiguously from the common splittings ($\sim$125~$\mu$eV at 6~T) of the transitions  measured in circular polarization, see Fig.~\ref{Fig2}. The parameters describing the local anisotropy (Eq.~\ref{eq:Hs}), the exchange strength (Eq.~\ref{eq:HX}) and the $A^0$ position anisotropy (Eq.~\ref{eq:Ha}) were manually  adjusted to reproduce  the most remarkable features of the experimental images : e.g. the zero-field anticrossing $\delta$, the FM-AFM exchange $\Delta$, the anticrossing $\Delta_a$ or the $\sigma^+/\sigma^-$ Zeeman splitting in Voigt configuration.  In contrast to our previous assumption in Ref.~\onlinecite{Kudelski07} and to the geometrical effects discussed in the case of II-VI QDs~\cite{Leger05a},  we did not include any heavy-hole light-hole mixing due to QD in-plane asymmetry to reproduce the anticrossing $\Delta_a$. Actually, the latter can be fully ascribed to the $A^0$ position anisotropy, since $\Delta_a\propto\rho\,\varepsilon_{\text{Mn-}h_2}\sin^2\theta_a$. The heavy-light hole splitting $\Delta_{HL}$ was  assumed to amount to a few 10~meV, so that the $h_2$ hole has a dominant heavy-hole character in agreement with the PL polarization in Fig.~\ref{Fig1}\cite{DeltaHL}. Besides, the strong circular polarization still measured in Voigt (up to 90\%, see Fig.~\ref{Fig2}(b)) indicates that the $h_2$ hole spin states  $|\Uparrow\rangle$ and $|\Downarrow\rangle$ are essentially not mixed by the magnetic field, in contrast to undoped InAs QD's where an effective transverse $g$-factor $g_{h_\perp}\!\sim$0.3 is found\cite{Bayer-PRB65,PRL-Xu07}. We ascribe this effect to the strong FM-AFM exchange $\Delta$ which inhibits this coupling for magnetic fields as long as  $|B_x|<\Delta/(g_{h_\perp}\mu_B)\sim$50~T. We still have  to  explain the splitting between the $\sigma^+$ and $\sigma^-$ components, namely between the $|\Uparrow,\pm1\rangle$ and  $|\Downarrow,\pm1\rangle$  states. This effect which is well reproduced by our model (see Fig.~\ref{Fig2}(c)) results from the  local anisotropy of $A^0$ which yields a finite in-plane spin projection  of the  $|\pm1\rangle$ states. The latter experience therefore  a Zeeman effect in a transverse magnetic field draging along the almost pure hole spin states $\Uparrow$ and $\Downarrow$. The resulting splitting  $\Delta_{\Uparrow,\Downarrow}$  of the  FM and AFM $h_2$-$A^0$ levels calculated within  second order perturbation theory reads\cite{note1} :\begin{equation}
\Delta_{\Uparrow,\Downarrow}=\frac{7 g_{Mn} -3 g_{h_1}}{5}  \frac{\cos\phi_s \, \sin 2\theta_s\, \left(\epsilon_\perp+\epsilon_\parallel\right)}{6\,\varepsilon_{h_1\text{-}h_2} - 14\,\varepsilon_{\text{Mn-}h_2}}\mu_\text{B} B_x
\label{eq:VoigtSplit}\end{equation}
As logically expected,  the splitting $\Delta_{\Uparrow,\Downarrow}$    vanishes for a magnetic field perpendicular to $z'$ ($\phi_s=\pm\pi/2$) or  for $\theta_s$=0~or~$\pi/2$ when cylindrical symmetry is restored.   Experimentally, we indeed observed a  large variety of magneto-PL images in Voigt configuration when measuring different Mn-doped InAs QDs. QD2 shown in Fig.~\ref{Fig3} is  very similar to QD1 in Faraday configuration but exhibits a very distinctive behavior in Voigt. The  Voigt magneto-PL image is almost symmetrical between $\sigma^+$ and $\sigma^-$ polarization, which means that individual lines are now weakly circularly polarized. Such a situation is expected  when  the main strain  axis $z'$ is almost orthogonal to the field direction. We could indeed reproduce QD2 magneto-PL images by taking $\phi_s$=93$^\circ$, whereas other fitting parameters were found close to those of QD1. Note however that  the anisotropic  exchange parameter $\rho$ is almost twice that for QD1. This was found necessary to reproduce a third zero-field doublet (denoted $O'$ in Fig.~\ref{Fig3}(a)) with the same  $\delta$ splitting as the FM and AFM doublets. This doublet corresponds  to the transition from the  two e-$A^0$  lower states (split by $\delta$ in zero field, see Fig.~\ref{Fig1}(c)) to the $J_z=0$ $h_2$-$A^0$ level. This normally forbidden transition becomes permitted here thanks to the strong anisotropic exchange $\propto\rho$.  Remarkably, the relative position of these three doublets reflects thus directly the level fan chart of $h_2$-$A^0$ in zero field.\\
\indent Figure \ref{Fig4} illustrates another interesting case (QD3), where the strain anisotropy experienced by $h_1$ is no longer dominated by a strong uniaxial strain $\epsilon_\parallel$ along $z'$. As shown in Fig.~\ref{Fig4}(a), the zero-field spectrum of QD3 strongly deviates from our usual observation of Mn-doped QDs. Up to 9 lines are visible, but   no doublet structures can  be clearly perceived. By performing the auto-correlation of these normalized peaks, we found that these lines perfectly reflect all 3$\times$3 possible transitions from $e$-$A^0$ to $h_2$-$A^0$ as indicated by  $i\,$\tiny$\blacktriangleright$\normalsize$j$ labels ($i,j=1,2,3$) in the figure. This is evidenced by the horizontal bars in Fig.\ref{Fig4}(a) which show that all  2-level splittings (in the initial or final  shells) appear three times in the measured spectrum. The  images in Voigt and Faraday showing both many anticrossings and non-linear field-dependencies confirm that these lines originate from the same Mn-doped QD, even though the cross-like pattern usually observed  is now hardly perceptible. Quite remarkably, our model  enables us to reproduce  still fairly well the experimental magneto-PL images. We essentially had to reduce the strength of $\epsilon_\parallel$  (see Tab.~\ref{ParamTable}) while keeping the other parameters close to those of QD1. This indeed leads to  eigenstates with  $A^0$ angular momentum  very different in the initial ($e$-$A^0$)  and final ($h_2$-$A^0$) shells, so that all transitions become partially allowed. In Voigt configuration, we still observe, like for QD1, a marked field-induced splitting due to the transverse Zeeman  effect of the $A^0$ complex. However, the circular polarization of the transitions is now much weaker ($<\sim$30\%) than for QD1. This is due to the mixing of the $h_2$ spin states  by the anisotropic exchange $\hat{\mathcal{H}}_a$, which is appreciably  enhanced with respect to QD1 because of the smaller strain-induced splittings of $A^0$ levels.\\
\indent Our experimental observations and numerical simulations indicate that so far we have  likely studied  Mn-doped InAs QDs with strongly uncentered Mn atom. This  sounds compatible with the fundamental issue raised in Ref.~\onlinecite{PRB78-Chutia} where a singlet configuration in a common S-like orbital is  predicted for the  $h_1$-$h_2$ ground state  when the Mn atom lies exactly at the QD center. Our model  describes instead two holes occupying distinct orbitals at different positions with a finite overlap between them giving rise to the AFM coupling  as for electrons in the H$_2$ molecule~\cite{Fazekas}.\\
\indent In conclusion, the  $\mu$-PL investigation in both a longitudinal and transverse magnetic field of  individual singly Mn-doped InAs QDs reveals remarkable new insights into the spin interactions between carriers and a Mn impurity in a III-V matrix. The explicit anisotropic part to the $p$-$d$  exchange~\cite{PRB-Bhattacharjee07} due to the Mn position with respect to the QD center explains better than the QD geometrical anisotropy~\cite{Leger05a}  certain forbidden transitions and  anti-crossings observed in Faraday  configuration. More spectacular is still the  conservation of circularly-polarized selection rules along with Zeeman splitting in a transverse magnetic field. This non-intuitive result  is remarkably   well interpreted  by considering  pure heavy-hole states in the quantum dot and  local potential anisotropy experienced by the acceptor level bound to the Mn impurity. The latter plays a crucial role to explain the dependence on the in-plane (azimuthal)  magnetic field angle, as well as the effective optical selection rules. Our results which validate the picture of a Mn impurity keeping a tightly bound hole in spite of the QD strain and  composition~\cite{PRB78-Chutia}, opens the way toward resonant experiments similar to those recently achieved in  semiconductor QD molecule~\cite{PRL-Kim08} in order to optically prepare and read out a single spin.\\

\begin{acknowledgments} We acknowledge fruitful discussions with A.~K.~Bhattachargee and S.~Chutia. This work was partly supported by the R\'{e}gion Ile-de-France and ANR contracts BOITQUANT and MOMES.
\end{acknowledgments}

\bibliographystyle{apsrev}

\begin{thebibliography}{25}
\bibitem[{Del()}]{DeltaHL}
\bibinfo{note}{This parameter cannot be determined with precision because it
  has basically no influence on the simulated spectra, as soon as it guarantees
  a clear sepraration between the heavy and light holes, namely above
  $\sim$10~meV.}
\bibitem[{not()}]{note1}
\bibinfo{note}{To simplify the general expression we took $\psi_s=0$.}
\end{thebibliography}

\end{document}